\documentclass[aps,prc,twocolumn,superscriptaddress,floatfix,showpacs,nofootinbib]{revtex4-2}
\usepackage{amssymb}
\usepackage{amsmath,bm}
\usepackage{newtxtext,newtxmath,booktabs,siunitx}
\usepackage[normalem]{ulem}
\usepackage[dvips]{color}
\usepackage{booktabs}
\usepackage{braket}
\usepackage{graphicx}
\usepackage{float}
\usepackage{multirow}
\usepackage{lineno}

\usepackage[normalem]{ulem}

\usepackage[usenames, dvipsnames]{xcolor}
\usepackage{tensor}
\usepackage[utf8]{inputenc}
\usepackage{subfigure}

\newcommand{\ac}{{\rm ac}_{2}\{3\}}

\newcommand{\pt}{p_{\rm T}}
\newcommand {\cosDPsi}  {\cos4(\Psi_4-\Psi_2)}
\newcommand{\evtcor}{\mean{\cos{4(\Psi_{4}-\Psi_{2})}}}
\newcommand{\rhovtpt}{\rho(v_{2}^{2},[\pt])}
\newcommand {\vtt}      {v_{2}\{2\}}

\newcommand {\vtf}      {v_{2}\{4\}}

\newcommand{\snn}{\sqrt{s_{\rm NN}}}

\newcommand{\U}{$^{238}$U}

\newcommand{\Au}{$^{197}$Au}

\newcommand {\mean}[1]  {\langle #1\rangle}

\begin{document}

\title{Systematic investigation of the nuclear multiple deformations in U+U collisions with A Multi-Phase Transport model}

\author{Zaining Wang}
\affiliation{Key Laboratory of Nuclear Physics and Ion-beam Application (MOE), Institute of Modern Physics, Fudan University, Shanghai 200433, China}
\affiliation{Shanghai Research Center for Theoretical Nuclear Physics, NSFC and Fudan University, Shanghai 200438, China}

\author{Jinhui Chen}
\email{chenjinhui@fudan.edu.cn}
\affiliation{Key Laboratory of Nuclear Physics and Ion-beam Application (MOE), Institute of Modern Physics, Fudan University, Shanghai 200433, China}
\affiliation{Shanghai Research Center for Theoretical Nuclear Physics, NSFC and Fudan University, Shanghai 200438, China}

\author{Hao-jie Xu}
\email{haojiexu@zjhu.edu.cn}
\affiliation{School of Science, Huzhou University, Huzhou, Zhejiang 313000, China}
\affiliation{Strong-Coupling Physics International Research Laboratory (SPiRL), Huzhou University, Huzhou, Zhejiang 313000, China}

\author{Jie Zhao}
\email{jie\_zhao@fudan.edu.cn}
\affiliation{Key Laboratory of Nuclear Physics and Ion-beam Application (MOE), Institute of Modern Physics, Fudan University, Shanghai 200433, China}
\affiliation{Shanghai Research Center for Theoretical Nuclear Physics, NSFC and Fudan University, Shanghai 200438, China}

\begin{abstract}
Relativistic heavy ion collisions provide a unique opportunity to study the shape of colliding nuclei, even up to higher-order multiple deformations. In this work, several observables that are sensitive to quadrupole and hexadecapole deformations of Uranium-238 in relativistic U+U collisions have been systematically investigated with A Multi-Phase Transport model. We find that the flow harmonic $v_{2}$, the $v_{2}$ and mean transverse momentum correlation, and the three-particle asymmetry cumulant $\ac$ are sensitive to nuclear quadrupole deformation, while $\ac$ and nonlinear response coefficient $\chi_{4,22}$ are sensitive to nuclear hexadecapole deformation. Our results from transport model studies are in qualitative agreement with previous hydrodynamic studies. The results indicate that the uncertainties of the hexadecapole deformation of Uranium  on the quadrupole deformation determination can be reduced by the abundance of correlation observables provided by the relativistic heavy ion collisions. 

\end{abstract}

\date{\today}
\maketitle

\section{Introduction}
A deconfined quantum chromodynamics (QCD) medium, the so-called quark-gluon plasma (QGP), is believed to be formed within a few yoctoseconds ($10^{-24} s$) in relativistic heavy ion collisions (HIC) at BNL's Relativistic Heavy Ion Collider (RHIC) and CERN's Large Hadron Collider (LHC)~\cite{Shuryak:1980tp,Adams:2005dq,Adcox:2004mh,ALICE:2010suc,Chen:2024zwk}. In HIC, the QCD strong interactions drive the initial geometry asymmetry of the QGP medium into an anisotropic distribution of final state hadrons in momentum space. 
Such an anisotropic momentum distribution can be described by flow harmonics $v_{n}$ using Fourier expansions of the azimuthal distribution of the particles~\cite{Ollitrault:1992bk,Kolb:2000fha}. The flow harmonics measured in relativistic HIC can be successfully described by relativistic hydrodynamics with the specific shear viscosity $\eta/s$ close to the quantum lower limit $1/4\pi$~\cite{Kovtun:2004de,Romatschke:2007mq,Song:2010mg,Shen:2020mgh,Wang:2022det,Deng:2023rfw}. In hydrodynamics, such a conversion is a consequence of the pressure gradient, and the hydrodynamic response of the lower order flow harmonics to the initial state is perfectly linear~\cite{Qiu:2011iv,Wei:2018xpm}, making it an ideal observable to trace back to spatial anisotropies in the initial state.  

The spatial anisotropies of the QGP medium are governed by the impact parameter (b) in non-head-on collisions~\cite{Ollitrault:1992bk}, while they are dominated by the deformation of the colliding nuclei in most-central collisions ($b\simeq 0$ fm)~\cite{Jia:2021tzt}. Several observables, such as the flow harmonics $v_{n}$, the mean transverse momentum fluctuations, and the flow harmonic correlations have been proposed to study the quadrupole ($\beta_{2}$) and octupole deformations ($\beta_{3}$) of the colliding nuclei in relativistic U+U collisions, Xe+Xe collisions, and isobar collisions~\cite{Heinz:2004ir,Luo:2007zf,Schenke:2020mbo,Giacalone:2021udy,Magdy:2022cvt,Zhao:2022uhl,STAR:2024eky,ATLAS:2022dov}. The advantage of such an unconventional way to study the nuclear shape is that the nuclear structure is imprinted into the initial stage of the produced QGP with instant snapshots (yoctoseconds), mostly decoupled from the subsequent bulk evolution~\cite{Zhang:2017xda,Li:2019kkh,Xu:2021uar,Jia:2021tzt,Wang:2023yis,YuanyuanWang:2024sgp}. 
Previous studies have focused on the lower-order nuclear multiple deformations, 
the heavy ion observables are expected to be less sensitive to the higher order deformations such as the hexadecapole deformations ($\beta_{4}$), as their effect would be overwhelmed by that of the lower-order multiple deformations.

However, recent studies of relativistic U+U and Au+Au collisions indicate the importance of hexadecapole deformation in relativistic HIC~\cite{STAR:2015mki,Ryssens:2023fkv,Xu:2024bdh}. The $v_{2}$ differences between the most central U+U and Au+Au collisions measured at the top RHIC energy indicate that hydrodynamic simulations require either larger $\beta_{2}$ for \Au\ or smaller $\beta_{2}$ for \U\ than those commonly used in Woods-Saxon-type nuclear densities to describe the data~\cite{Schenke:2020mbo,Giacalone:2021udy,Ryssens:2023fkv,Xu:2024bdh}. Such ambiguities cannot be avoided with the exact value of the ground state electrical transition rates $B(E2)=12.09 \pm 0.20$ e$^{2}$b$^2$ of Uranium measured by the low-energy experiment~\cite{Raman:2001nnq}, since the magnitude of $\beta_{2}$ depends strongly on $\beta_{4}$ for a given $B(E2)$~\cite{Ryssens:2023fkv,Ryssens:2014bqa}. Considering the uncertainty of $\beta_{4}$ in the low-energy experiment measurements and nuclear structure theory calculations, the observables proposed in relativistic HIC to extract the $\beta_{4}$ of Uranium turn out to be meaningful~\cite{Xu:2024bdh}.

In this study we consider a moderate uncertainty in the $\beta_{4}$ of Uranium~\cite{Bemis:1973zza,Zumbro:1984zz}, i.e. $\beta_{4}=0.1$ vs. $\beta_{4}=0$, such an uncertainty gives $\Delta\beta_{2}\simeq0.03$ for its quadrupole deformation uncertainty~\cite{Ryssens:2023fkv,Xu:2024bdh}. The question is whether this uncertainty can be systematically reduced or eliminated by the observables obtained by relativistic HIC. The answer seems to be positive for the hydrodynamic simulations~\cite{Xu:2024bdh}. However, the discrepancy between the $\beta_{2}$ of \U\ extracted from $v_{2}$ and other observables from hydrodynamic simulations~\cite{STAR:2024eky} makes us think carefully about the uncertainties introduced by the models. Therefore, in this work, we use a transport model (A Multiphase Transport model AMPT)~\cite{Lin:2004en,Wang:1991hta,Li:1995pra,Zhang:1997ej} to systematically investigate the observables that are sensitive to nuclear quadrupole and hexadecapole deformations in relativistic HIC. The AMPT model is generally considered to have a similar response to flow as hydrodynamics, while some other mechanisms, such as the  anisotropic parton escape mechanism~\cite{He:2015hfa}, are also proposed to be dramatically different from the hydrodynamic scenario.

The rest of the paper is organized as follows. Section~\ref{sec:model} gives a brief description of the AMPT model and the definition of the observables used in this work. Section~\ref{sec:results} discusses the effects of nuclear quadrupole and hexadecapole deformations on the flow-related observables in AMPT simulations. A summary is given in Sec.~\ref{sec:summary}.

\section{Model setup and analysis methods}
\label{sec:model}
\subsection{AMPT model}
We perform all the calculations of the observables within AMPT model~\cite{Lin:2004en,Wang:1991hta,Li:1995pra,Zhang:1997ej} to study the effects of different nuclear deformations on these observables in U+U collisions at $\sqrt{s_{\rm NN}}$ = 193 GeV and Au+Au collisions at $\sqrt{s_{\rm NN}}$ = 200 GeV. 
The AMPT model aims to apply the kinetic theory approach to describe the evolution of relativistic HIC as it contains four main components: the fluctuating initial conditions, partonic interactions, hadronization, and hadronic interactions~\cite{Lin:2004en}. 
It has since been widely used to simulate the evolution of the dense matter created in high-energy nuclear collisions. In particular, the string melting version of the AMPT model can well describe the anisotropic flows and particle correlations in collisions of $pp$, $pA$, or $AA$ systems at RHIC and LHC energies
~\cite{Bzdak:2014dia,Zhang:2020hww,Li:2021znq,Shen:2021pds,Zhang:2021ygs,Wang:2022fwq,Shao:2022eyd,Zhu:2022dlc,Zhao:2021bef,Shen:2022gtl}. 
In our study, we use the string melting version of AMPT,
and some of the key parameters are the Lund string fragmentation parameters $a_L$ = 0.5 and $b_L$ = 0.9 GeV$^{-2}$, the parton screening mass $\mu_{D}=3.2032$ fm$^{-1}$, and the strong coupling constant $\alpha_{S}=0.33$, corresponding to a total cross section $\sigma=1.5$ mb. 
Since we focus on the nuclear deformation effect in most central collisions in this study, the AMPT with $\sigma=1.5$ mb was found to give better predictions on $v_{2}$ and $v_{4}$ in most central Au+Au collisions at $\snn=200$ GeV than those with $\sigma=3$ mb~\cite{Nasim:2016rfv}.

The shape of the nucleus with a finite number of nucleons distributed with a density $\rho$ can be described by the Woods-Saxon distribution~\cite{Woods:1954zz}
\begin{eqnarray}
\rho(r,\theta) &=& \frac{\rho_0}{1+ {\rm exp}[(r-R)/a]},\\
R &=& R_0(1+\beta_2 Y_2^0(\theta)+\beta_4 Y_4^0(\theta)),
\end{eqnarray}
where $R_{0}$ is the radius parameter, $a$ is the diffuseness parameter, $\beta_2$ and $\beta_4$ are the deformation parameters, and $Y_2^0$, $Y_4^0$ are spherical harmonics.
To study the uncertainties of $\beta_{2}$ from $\beta_{4}$, we use three sets of parameters for \U\ with $\Delta\beta_{2}=0.035$ and $\Delta\beta_{4}=0.1$, keeping other parameters same~\cite{Raman:2001nnq,Ryssens:2023fkv,DeJager:1987qc,Pritychenko:2013gwa}. 
The deformation parameters are listed in Tab~\ref{Tab1}. Calculations of 0-60\% centrality are performed and charged particles are selected with $|\eta|<2.0$ and $\pt>0.2$ GeV/c.
To focus on the most central collisions, 3 million events with $\mathrm{b_{max}}=20$ fm for each case and 1 million events with $\mathrm{b_{max}} = 4.74$ fm for U+U collisions are generated. The collision configurations are generated randomly with Euler rotations.
\begin{table}
\centering
\caption{Deformation parameters applied in the AMPT model. Here we use Au + Au collisions as a reference in the computation of observables.}

\begin{tabular}{lcccrr}
\hline
\multicolumn{2}{c}{Species} &\hspace{0.2cm} $R_0$ (fm) \hspace{0.2cm} & \hspace{0.2cm} $a$ (fm) \hspace    {0.2cm} & \hspace{0.5cm} $\beta_{2}$ \hspace{0.1cm} & \hspace{0.7cm} $\beta_{4}$ \hspace{0.1cm}   \\
\hline  
\hline
\multirow{3}{*}{U} & a)  & 6.80  & 0.615   & 0.282  &  0.00 \\
& b)  & 6.80  & 0.615   & 0.247  &  0.10 \\
& c)  & 6.80  & 0.615   & 0.282  &  0.10 \\
\hline
\multicolumn{2}{c}{Au}  & 6.38  & 0.535   & -0.131 & 0.00 \\
\hline
\end{tabular}
\label{Tab1}
\end{table}

In this work, the flow harmonics $v_{n}$, the mean transverse momentum fluctuation, the Pearson correlation coefficient $\rhovtpt$ between $v_{n}$ and mean transverse momentum, three-particle asymmetry cumulant $\ac$, and the nonlinear response coefficient $\chi_{4,22}$ between $v_{4}$ and $v_{2}$ in relativistic U+U collisions at $\snn=193$ GeV have been systematically investigated.
We also use Au + Au collisions at $\snn=200$ GeV as a reference in the computation of observables, with their Woods-Saxon density parameters set to the commonly used values~\cite{Loizides:2017ack}. The results are shown as a ratio from U+U collisions to Au+Au collisions
\begin{equation}
R(X)\equiv \frac{X_{\rm UU}}{X_{\rm AuAu}}.
\end{equation}
Before the discussion of the results, some definitions of these observables are given below.

\subsection{Observables}
The azimuthal dependence of the final particle distribution can be written as
\begin{eqnarray}
\frac{2\pi}{N}\frac{dN}{d\phi}=1+\sum\limits_{n=1}^{\infty}2v_n\cos n(\phi-\Psi_n).
\end{eqnarray}
Here $v_n$ are the flow harmonics, $\phi$ is the azimuthal angle of the momentum of the outgoing particles, and $\Psi_n$ is the event plane angle~\cite{Poskanzer:1998yz} defined as $\langle e^{in\phi}\rangle=v_ne^{in\Psi_n}$, where $\langle..\rangle$ is the average in a given event. 
The $v_{n}(\Psi_{n})$ can be calculated with the two-particle correlation method $v_{n}\{2\}$. In this work, the standard Q-cumulant method is used to calculate the flow observables~\cite{Bilandzic:2010jr}.

For high order flow harmonics $v_{n}$, it can be calculated with respect to the lower-order event plane angle, not just the same order one $\Psi_{n}$~\cite{Poskanzer:1998yz}. For example, the $v_{4}(\Psi_{2})$ calculated with respect to the second-order event plane angular $\Psi_{2}$ is directly related to the nonlinear part of $v_{4}$~\cite{Yan:2015jma} 
\begin{eqnarray}
v_{4}\{2\}&=&v_{4}(\Psi_{4})=v_{\rm 4L}+ v_{\rm 4NL},  \\
v_{\rm 4NL}  &=& v_{4}(\Psi_{2}) = \chi_{4,22}(v_2)^2,\label{con:V4}
\end{eqnarray}
where $v_{\rm 4L}$($v_{\rm 4NL}$) is the (non-)linear part of $v_4$, and $\chi_{4,22}$ is the nonlinear response coefficient.

The three-particle asymmetry cumulant, 
\begin{equation}
\ac \equiv \mean{\mean{3} _{2,2,-4}}=\langle v_{2}^{4}\rangle^{1/2} v_{4}\{\Psi_{2}\}\,,
\end{equation}
reflects the flow harmonic correlation between $v_{2}$ and $v_{4}$~\cite{ATLAS:2014qaj,Yan:2015jma, Jia:2017hbm, Zhao:2022uhl}.
Here $\langle v_{2}^{4}\rangle =\langle\langle 4\rangle_{2,2,-2,-2}\rangle=2\vtt^{4}-\vtf^{4}$ denotes the four-particle moment, 
with the multi-particle azimuthal moment~\cite{Bilandzic:2010jr, Bilandzic:2013kga} given by
$\mean{m}_{n_1,n_2,...,n_m} \equiv \mean{e^{i(n_{1}\varphi_{k_{1}}+n_{2}\varphi_{k_{2}}+...+n_{m}\varphi_{k_{m}})}}$,
where $\mean{\cdot\cdot}$ averages over all particles of interest (POI) in a given event, and an outer $\mean{\mean{\cdot\cdot}}$ denotes further average over an ensemble of events.
$\ac$ can be roughly written as~\cite{Yan:2015jma}
\begin{equation}
\ac = \mean{v_{2}^{2}v_{4}\cosDPsi}.
\label{eq:ac}
\end{equation}
The nonlinear response coefficient is then given by~\cite{Yan:2015jma,Xu:2024bdh},
 \begin{equation}
 \chi_{4,22} \equiv \frac{v_{4}\{\Psi_{2}\}}{\langle v_{2}^{4}\rangle^{1/2}} 
  = \frac{\ac}{\langle v_{2}^{4}\rangle}\,.
 \end{equation}
In hydrodynamic scenarios, $\ac$ and $\evtcor$ are sensitive to $\beta_2$ and $\beta_3$, 
while $\chi_{4,22}$ can be considered an ideal observable to probe $\beta_{4}$ as it is only sensitive to nuclear hexadecapole deformation ~\cite{Xu:2024bdh}.

The Pearson correlation coefficient $\rho(v_2^2,[\pt])$ between $v_2$ and the mean transverse momentum $[\pt]$ is found to have a strong sensitivity to the nuclear deformations~\cite{Giacalone:2020awm,Jia:2021wbq}, which can be calculated as
\begin{eqnarray}
\rho(v_n^2,[\pt]) = \frac{\mathrm{Cov}(v_n^2,[\pt])}{\sigma_{\pt}\sigma_{v_n^2}},
\end{eqnarray}
where $[\pt]$ is the averaged transverse momentum in a given event, Cov($v_n^2$,$[\pt]$) is the covariance between $v_n^2$ and $[\pt]$ and $\sigma$ is the standard deviation. The deviation of $v_n$ can be calculated by subtracting the four-particle correlation from the two-particle correlation. 
Besides $\rho_{2}$, the $[\pt]$ variance $\langle(\delta \pt)^2\rangle \equiv \sigma_{\pt}^{2}$ is also sensitive to nuclear deformation~\cite{Jia:2021qyu}, where $\delta \pt$ = $[\pt] - \langle [\pt] \rangle_{\mathrm{event}}$ is the fluctuation of $[\pt]$ in a given centrality range and $\langle...\rangle$ denotes the average over an ensemble of events.

\section{Results and discussions}
\label{sec:results}

\begin{figure}
\centering
\includegraphics[width=0.48\textwidth]{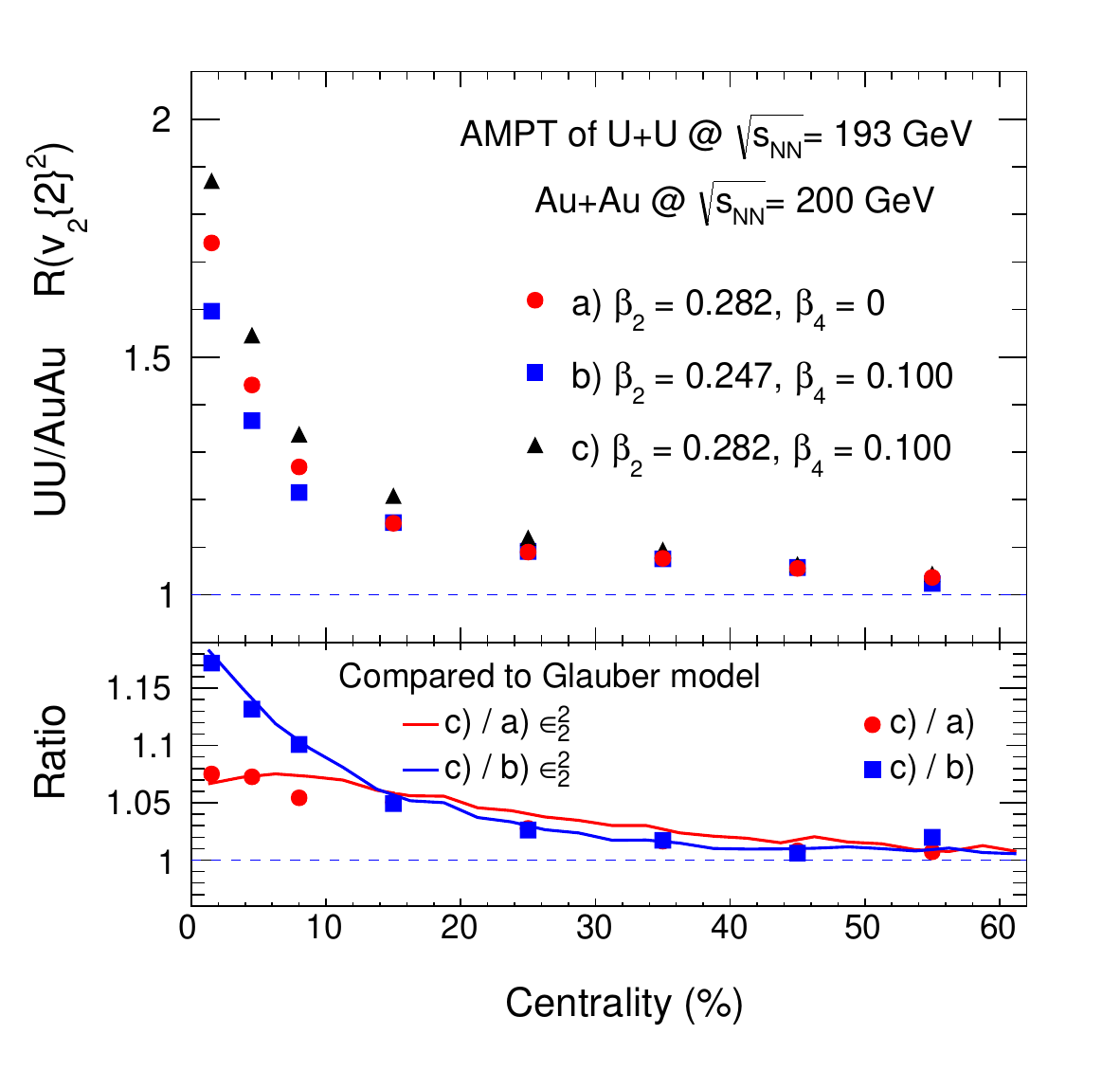}
\caption{(Color online) The ratios of the values of $v_{2}\{2\}^{2}$ from U+U collisions to Au+Au collisions in the 0-60\% centrality interval calculated from AMPT simulations. These cumulants are calculated by the standard Q-cumulant method using two-particle correlations. The bottom panel contains ratios of $v_{2}^{2}$ from the AMPT model and $\epsilon_{2}^{2}$ from the Monte Carlo Glauber model with different deformation parameters listed in Tab.~\ref{Tab1}. }
\label{v2}
\end{figure} 

\begin{figure*}
\centering
\includegraphics[width=0.48\textwidth]{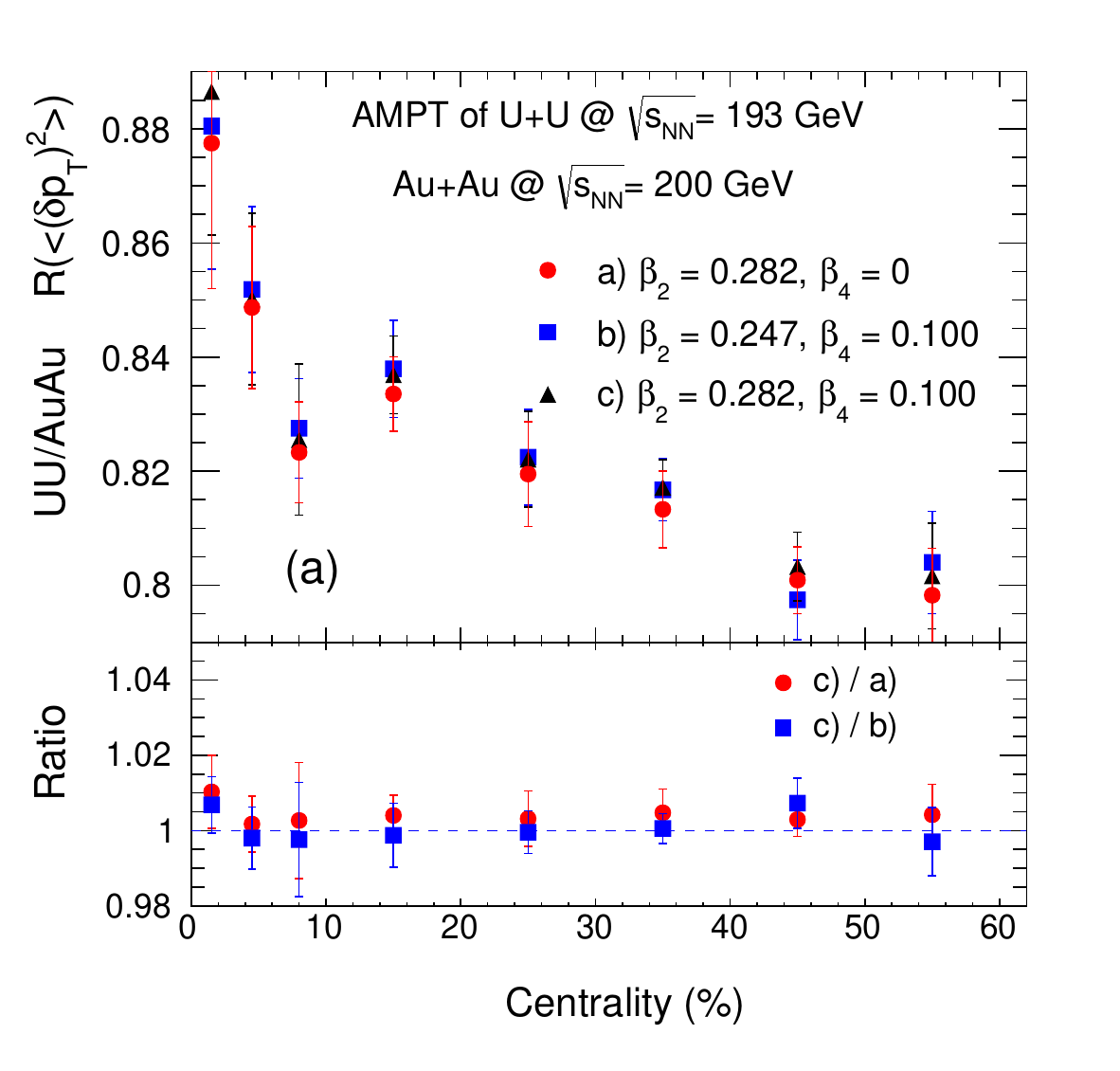}
\includegraphics[width=0.48\textwidth]{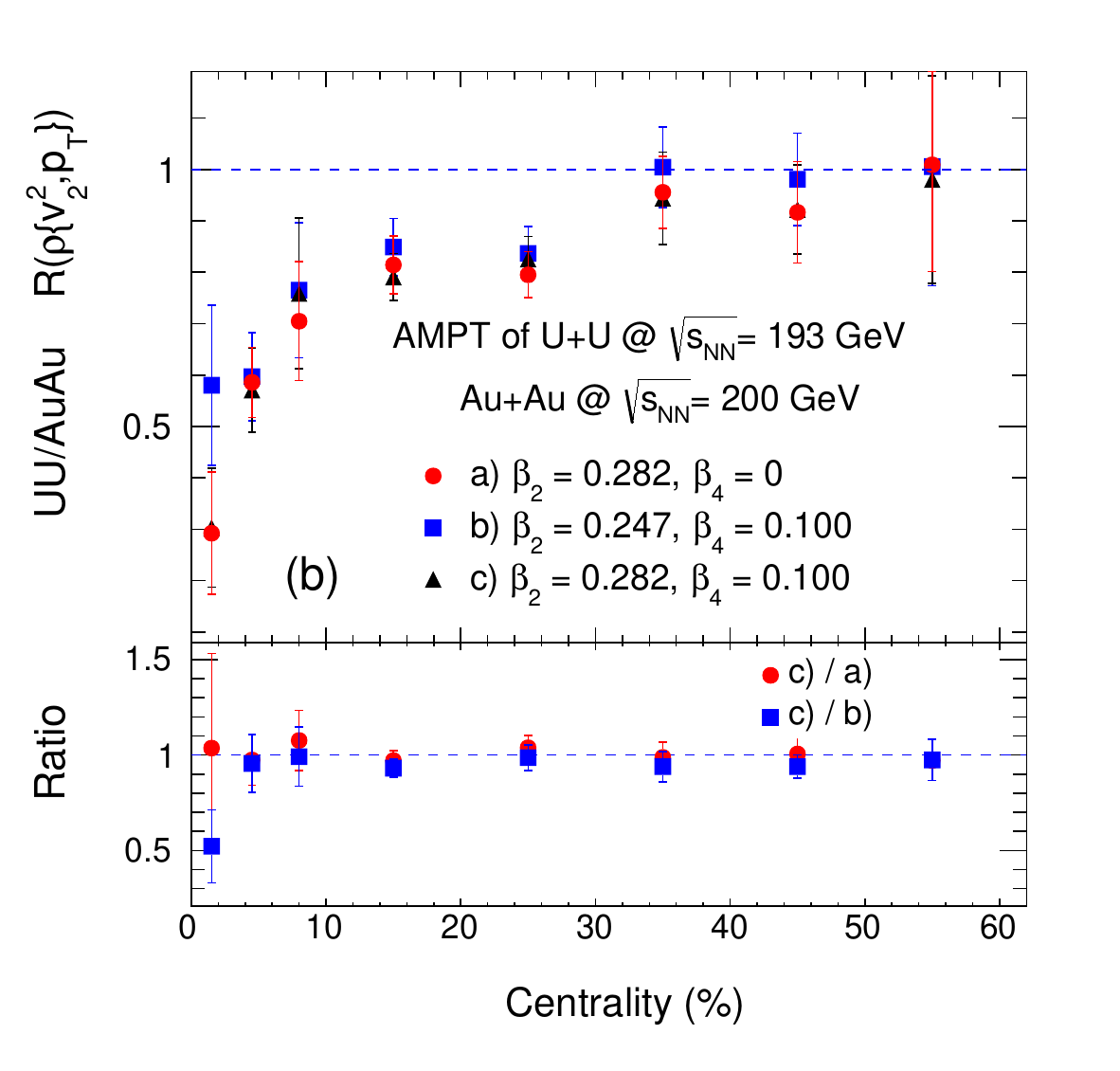}
\caption{(Color online) The ratios of the values of (a) the $[\pt]$ variance $\delta(\pt)^2$ and (b) the Pearson correlation coefficient $\rhovtpt$ of U+U collisions to Au+Au collisions calculated in the 0-60\% centrality interval from AMPT simulations. }
\label{pt}
\end{figure*} 

Figure~\ref{v2} shows the $v_{2}\{2\}^{2}$ difference between U+U collisions and Au+Au collisions with different $\beta_{2}$ and $\beta_{4}$ of Uranium. The deformation effect is highlighted in the lower panel with the double ratios, where the contributions from the reference Au+Au collisions are canceled out. 
The quadrupole deformation has a significant contribution to $v_{2}$ at most central collisions, and this feature has been found for decades~\cite{Heinz:2004ir}. We find that the $v_{2}\{2\}^{2}$ at most central collisions can be clearly distinguished with $\beta_{2}=0.282$ and $\beta_{2}=0.247$ from the AMPT simulations. The small $\beta_{2}$ of Uranium as a result of the suppression by $\beta_4$ have been used to explain the $R(v_{2}^{2})$ data measured at RHIC~\cite{Ryssens:2023fkv}, ignoring the uncertainties from the quadrupole deformation of gold. The effect of the hexadecapole deformation on $v_{2}$ is relatively small, which is consistent with previous studies~\cite{Xu:2024bdh,Jia:2021qyu}.

The AMPT model is known to have a similar response to the elliptic flow from medium expansion as hydrodynamics, i.e., $v_{2}\propto \epsilon_{2}$ holds for most of the centrality~\cite{Qiu:2011iv,Wei:2018xpm}.
 Therefore, the $R(v_{2}\{2\}^{2})$ can be well reproduced by the initial eccentricity ratio $R(\epsilon_{2}^{2})$, where $\epsilon_{n}e^{in\Phi_{n}}\equiv -\langle r^{n}e^{in\varphi}\rangle/\langle r^{n}\rangle $ with $(r,\varphi)$ being the position of the participant in the transverse area and $\Phi_{n}$ being the initial symmetry plane angle~\cite{Qiu:2011iv,Zhu:2016puf}. Here $\langle...\rangle$ denotes the averages over all participants. The double ratios of $\epsilon_{2}^{2}$ obtained by a Monte Carlo Glauber simulation ~\cite{Miller:2007ri,Loizides:2017ack} are also shown in the lower panel of Fig.~\ref{v2}. 
We confirm that the initial predictor $\epsilon_{2}$ works well for all the centrality, as the results from AMPT and Monte Carlo Glauber simulations are almost overlapped.  Due to the consistency among the macroscopic hydrodynamic model, the microscopic transport AMPT model, as well as the pure initial geometry Monte Carlo Glauber model, the nuclear quadrupole deformation extracted from elliptic flow ratio $R(v_{2}\{2\}^{2})$ between U+U collisions and Au+Au collisions is expected to have small model uncertainties. One ambiguity may be due to the hexadecapole deformation of the Uranium we have mentioned in the introduction~\cite{Ryssens:2023fkv}, which cannot be well controlled by the elliptic flow data~\cite{Xu:2024bdh}.  

Besides the elliptic flow $v_{2}$, the mean transverse momentum fluctuation $\langle(\delta \pt)^{2}\rangle$ and the Pearson correlation coefficient $\rho(v_{2}^{2},\pt)$ have also been used to extract the quadrupole deformation of the colliding nuclei~\cite{Giacalone:2020awm,Jia:2021qyu,STAR:2024eky}. The AMPT model is known to be a poor description of the mean transverse momentum and its fluctuation data, and some progress has been made on this issue~\cite{Lin:2021mdn,Zhang:2021vvp, Zhao:2024feh}. Although with such drawback, it is worth investigating the deformation effect on  $\langle(\delta \pt)^{2}\rangle$ and $\rho(v_{2}^{2},\pt)$ differences between U+U collision and Au+Au collisions with AMPT simulations. The results are shown in Fig.~\ref{pt}. The deformation effect of both finite $\beta_{2}$ difference and $\beta_{4}$ difference on the $[\pt]$ variance is invisible, see Fig.~\ref{pt}(a), in contrast to the conclusion from hydrodynamics~\cite{STAR:2024eky}. Despite the absolute magnitudes, the $R(\langle(\delta \pt)^{2}\rangle)$ shows a decreasing trend as a function of centrality, although much weaker, which is consistent with experimental data and hydrodynamic simulations~\cite{STAR:2024eky}.

\begin{figure}
\centering
\includegraphics[width=0.48\textwidth]{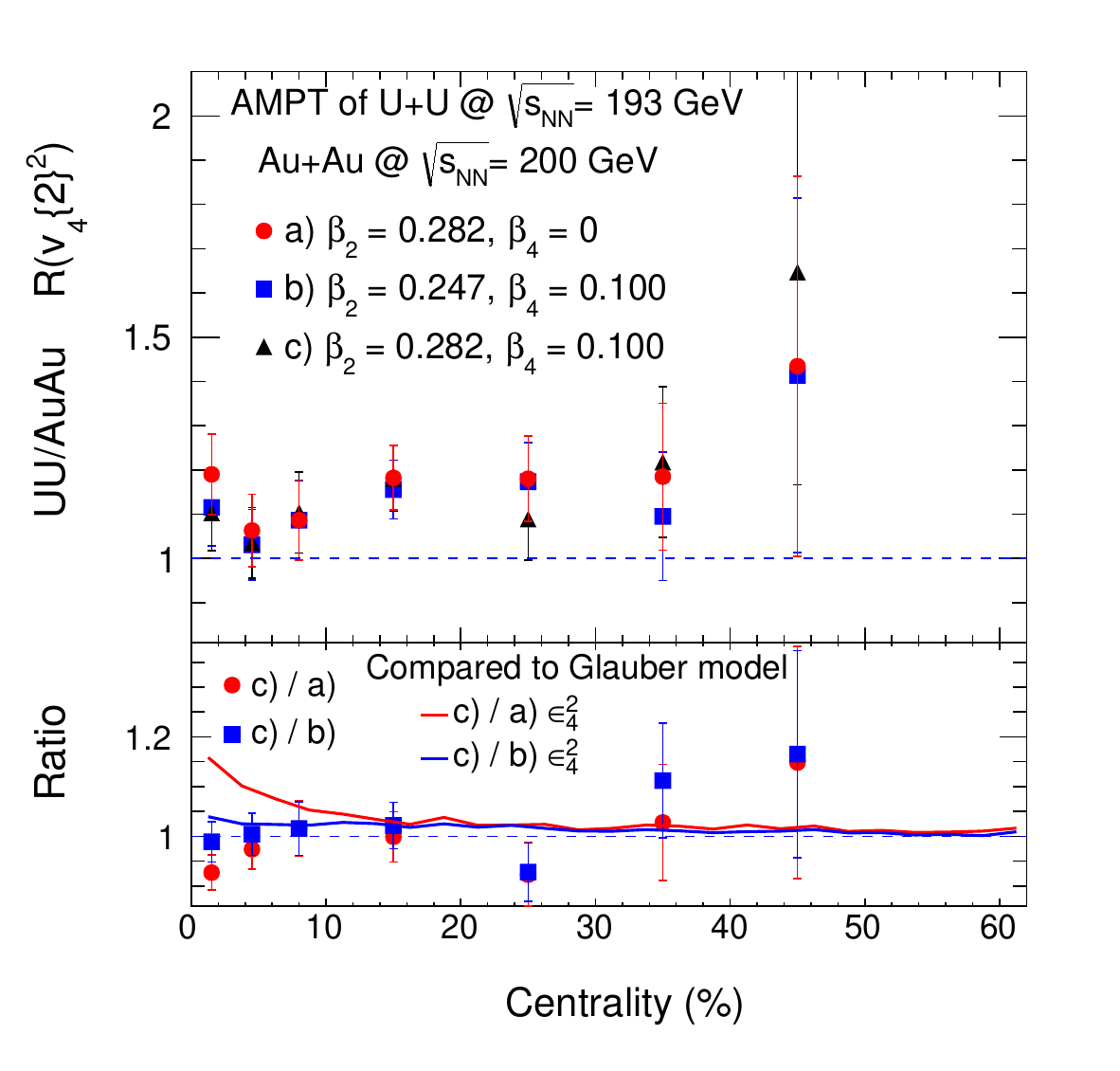}

\caption{(Color online) Similar to Fig.~\ref{v2}, but for the hexadecapole flow $v_{4}\{2\}^{2}$ and $\epsilon_{4}^{2}$. }
\label{v4}
\end{figure} 

\begin{figure*}
\centering
\includegraphics[width=0.48\textwidth]{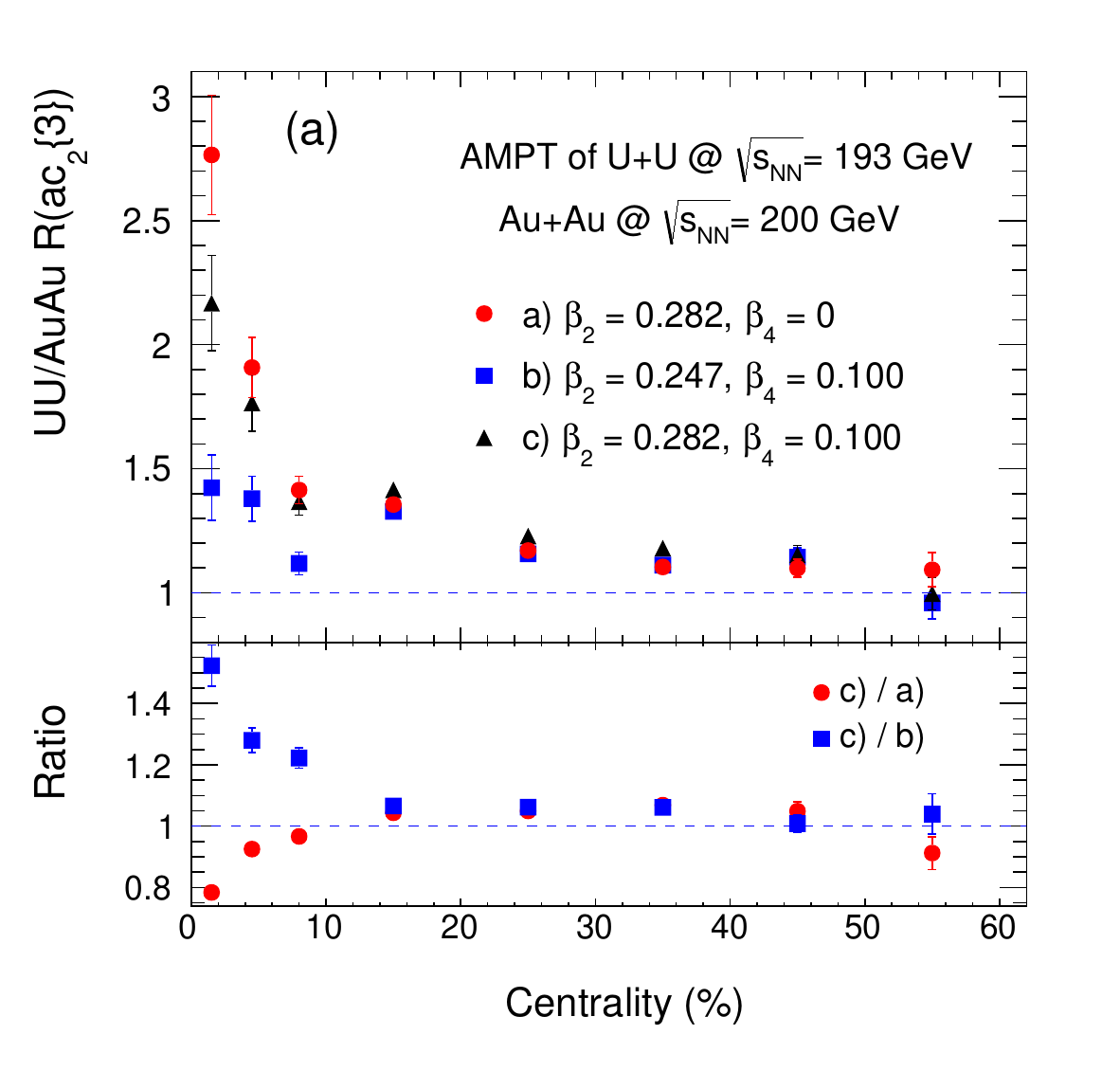}
\includegraphics[width=0.48\textwidth]{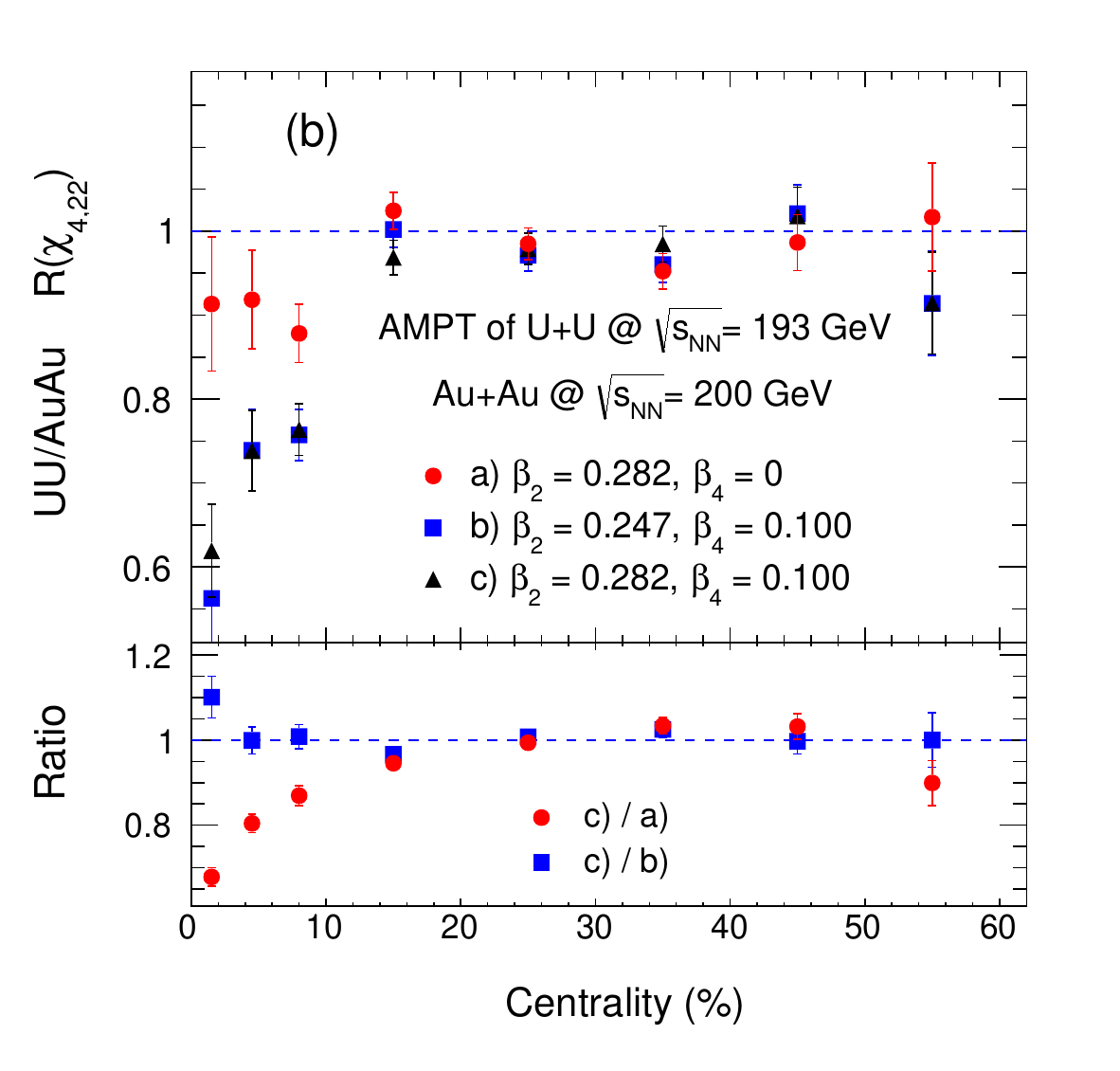}
\caption{(Color online) Similar to Fig.~\ref{pt}, but for (a) the three-particle asymmetry cumulant $\ac$ and (b) the nonlinear response coefficient $\chi_{4,22}$. }
\label{ac}
\end{figure*}

Due to the limited statistics in our study, we cannot make a firm conclusion on the effect of nuclear deformation on $\rho(v_{2}^{2},\pt)$, especially for the hexadecapole deformation. It appears that the quadrupole deformation gives large negative contribution to $\rho(v_{2}^{2},\pt)$ for most central collisions, and the hexadecapole deformation effect is negligible for all the centrality. 
These conclusions can be readily confirmed with more statistics.
Moreover, the $v_{2}$-$\pt$ correlations are also found to be sensitive to triaxial deformation~\cite{Bally:2021qys}. Due to the drawback of the current AMPT version, we propose such an AMPT study in the future. However, we note that the centrality-dependent trends of  $R(\rho(v_{2}^{2},\pt))$ are in qualitative agreement with the experimental data and hydrodynamic simulations~\cite{STAR:2024eky}. The deformation effect on $\langle(\delta \pt)^{2}\rangle$ and $\rho(v_{2}^{2},\pt)$ has been studied in previous study using AMPT model with an even stronger partonic cross section $\sigma=3$ mb~\cite{Jia:2021qyu}. We expect that these results can be used to provide some insights into the improvement of the AMPT model~\cite{Zhang:2021vvp,Lin:2021mdn}.

The observables discussed above are nearly insensitive to hexadecapole deformation. The effect of $\beta_{4}$ on heavy ion observables is rarely studied, as it is typically overwhelmed by that of quadrupole deformations. Recent studies suggest that $\beta_{4}$ may play a very important role in explaining the elliptic flow difference between U+U collisions and Au+Au collisions~\cite{Ryssens:2023fkv}. Based on hydrodynamic simulations, several observables have been proposed to extract the value of $\beta_{4}$ in relativistic HIC~\cite{Xu:2024bdh}. It is therefore worthwhile to test the sensitivity of these observables to $\beta_{4}$ with a transport model simulation. In addition, the hydrodynamic response of the AMPT model to higher-order flow harmonics has not been extensively studied. Therefore, we will focus on these $\beta_{4}$-sensitive observables that are related to the high-order flow harmonics in the rest of this paper.

Figure~\ref{v4} presents the effect of $\beta_{2}$ and $\beta_{4}$ on the hexadecapole flow $v_{4}$. With the $\beta_{2}$ and $\beta_{4}$ differences discussed in this study, the quadrupole and hexadecapole deformation effects on $v_{4}$ are negligibly small. This is not the case for the related initial predictor $\epsilon_{4}$. The $\epsilon_{4}$ differences calculated with Monte Carlo Glauber simulations using the same nuclear deformation differences are also shown in the lower panel of Fig.~\ref{v4}. We find that the enhancement of $\epsilon_{4}$ by $\beta_{4}$ is obvious. The evolution driven by strong interactions not only distorts this enhancement but even gives a negative contribution to the final hexadecapole flow. This is similar to the results from hydrodynamic simulations~\cite{Xu:2024bdh}. The results indicate that the response of high order flow such as $v_{4}$ in AMPT simulation is also non-diagonal and nonlinear~\cite{Qiu:2011iv}.

The $v_{4}$ is typically divided into the linear part $v_{\rm 4L}$ and the nonlinear part $v_{\rm 4NL}$~\cite{Yan:2015jma}. The three-particle asymmetry cumulant $\ac$ is directly related to $v_{\rm 4NL}$, but more statistical friendly~\cite{Zhang:2018lls,Zhao:2022uhl}. The effect of nuclear deformation on $\ac$ shown in Fig.~\ref{ac}(a) is quite obvious. The $\ac$ increase with $\beta_{2}$, but decreases with $\beta_{4}$, in very good agreement with previous hydrodynamic simulations~\cite{Xu:2024bdh}. From Eq.~\ref{eq:ac}, one would expect that the $\beta_{4}$ dependence is roughly from the flow angle correlation $\cos4(\Psi_{2}-\Psi_{4})$, as $v_{2}$ and $v_{4}$ are almost insensitive to $\beta_{4}$ discussed before. This is confirmed by our AMPT simulations (not shown). The competition between $\beta_{2}$ and $\beta_{4}$ on $\ac$ suggests that the extraction of $\beta_{2}$ differences from $\ac$ in relativistic HIC should be re-examined with the potential hexadecapole deformation of the colliding nuclei~\cite{Zhao:2022uhl}. This would be an interesting topic for future work.  

Previous studies with relativistic hydrodynamic simulations indicate that the nonlinear response coefficient $\chi_{4,22}$ can be considered as an ideal observable to extract the parameter $\beta_{4}$ in relativistic HIC~\cite{Xu:2024bdh}. The effect of $\beta_{2}$ and $\beta_{4}$ on the nonlinear response coefficient $\chi_{4,22}$ from the AMPT simulation is shown in Fig.~\ref{ac}(b). The $\chi_{4,22}$ is significantly suppressed by the finite $\beta_{4}$, while the contributions from the finite $\beta_{2}$ differences are reasonably small, in quantitative agreement with the predictions from hydrodynamic simulations~\cite{Xu:2024bdh}. The centrality dependence of $R(\chi_{4,22})$ from AMPT simulations converge to units for non-central collisions, suggesting that differences in system size and/or nuclear quadrupole deformations have a negligible effect on $\chi_{4,22}$. The results indicate that the AMPT has a similar nonlinear response to high order flow harmonic $v_{4}$ as hydrodynamics~\cite{Xu:2024bdh}, making $\chi_{4,22}$ as an ideal observable to probe nuclear hexadecapole deformations with small model uncertainties. Quantitatively, the effect of $\beta_{4}$ on $\chi_{4,22}$ is slightly ($\sim20\%$) weaker than the hydrodynamic response shown in Ref.~\cite{Xu:2024bdh}. These model uncertainties are crucial for the accurate extraction of $\beta_{4}$ from relativistic HIC. Therefore, more precise investigations of the differences in the linear and nonlinear response of the hexadecapole flow between the transport model and the hydrodynamic model are required. We leave these investigations in the forthcoming paper but focus on a more ideal platform -- relativistic isobar collisions~\cite{Deng:2016knn,Xu:2017zcn,Xu:2021vpn,Zhang:2021kxj,STAR:2021mii}.

In this paper, we focus mainly on the prediction differences between AMPT simulations and previous hydrodynamic simulations.
We note that the differences in system sizes between U+U and Au+Au collisions may also have residual contributions to model uncertainties, despite the potential differences in response mechanisms between the transport and hydrodynamic models. One might expect that the ratios of the two systems would cancel out the uncertainties from the bulk evolution. This may be true for relativistic isobar collisions, but not for the case discussed in this study, since \U\ and \Au\ have almost $\sim20\%$ relative differences in mass number. The contributions from the non-flow would make the situation even worse~\cite{Borghini:2000cm,Wang:2008gp}. Therefore, although the hydrodynamic model is considered the "standard model" for high-energy HIC, a comprehensive understanding of the underlying physical differences between AMPT and hydrodynamics is crucial for the study of higher-order flow harmonics, especially for the proposal of accurate extraction of nuclear deformation parameters in relativistic HIC.

\section{Summary}
\label{sec:summary}
Anisotropic flows and their correlations in most central collisions are key observables to study the nuclear deformation effect in relativistic HIC. In this work, the flow harmonics $\vtt$ and the related correlation observables such as the Pearson correlation coefficient $\rhovtpt$ between the anisotropic flow and the mean transverse momentum, the three-particle asymmetry cumulant $\ac$, and the nonlinear response coefficient $\chi_{4,22}$ between $v_{4}$ and $v_{2}$ in relativistic U+U collisions at $\snn=193$ GeV have been systematically studied with A Multiphase Transport model. We found that the former two observables $\vtt$ and $\rhovtpt$ are sensitive to quadrupole deformation, while the latter one $\chi_{4,22}$ is mostly sensitive to hexadecapole deformation. The $\ac$ is sensitive to both quadrupole and hexadecapole deformations. These features are in qualitative agreement with the results from hydrodynamic simulations. Due to the incorrect response of the radial flow, the $[\pt]$ variance in AMPT simulations is insensitive to nuclear deformation, in contrast to the hydrodynamic scenario. 
Our results suggest that $v_{2}$ ($\chi_{4,22}$) can be considered as an ideal observable to extract the nuclear quadrupole (hexadecapole) deformation parameter with small model uncertainties, although these uncertainties need to be investigated in detail for more precise constraints.
The results indicate that the uncertainty of the $\beta_{2}$ measurement introduced by the unknown hexadecapole deformation can be largely reduced by more deformation-sensitive observables in relativistic HIC.

\section*{Acknowledgements.}
We thank Zi-Wei Lin for providing the AMPT code and for fruitful discussions. This work was supported in part by the National Key Research and Development Program of China under Contract No. 2022YFA1604900, by the National Natural Science Foundation of China (NSFC) under Contract No. 12025501, No. 12035006, No. 12075085, No. 12147101, No. 12275082, No. 12275053,  by the Strategic Priority Research Program of Chinese Academy of Sciences under Grant No. XDB34030000, and by the Guangdong Major Project of Basic and Applied Basic Research No. 2020B0301030008.

\bibliography{myref}

\end{document}